\documentstyle[epsfig]{mn}{}

\begin{document}

\def\mpc{h^{-1} {\rm{Mpc}}}
\def\up{h^{-3} {\rm{Mpc^3}}}
\def\uk{h {\rm{Mpc^{-1}}}}
\def\lsim{\mathrel{\hbox{\rlap{\hbox{\lower4pt\hbox{$\sim$}}}\hbox{$<$}}}}
\def\gsim{\mathrel{\hbox{\rlap{\hbox{\lower4pt\hbox{$\sim$}}}\hbox{$>$}}}}
\def\kms {\rm{km~s^{-1}}}
\def\apj {ApJ}
\def\aj {AJ}
\def\aa {A \& A}
\def\mnras {MNRAS}

\title{Galaxy groups in the 2dF galaxy redshift survey: \\
A Compactness Analysis of Groups}
\author[Zandivarez et al.]{
\parbox[t]{\textwidth}{
A. Zandivarez$^{1}$, M.J.L. Dom\'{\i}nguez$^{1}$, C.J. Ragone$^{1}$, 
H. Muriel$^{1,2}$ \& H.J. Mart\'{\i}nez$^{1}$
}
\vspace*{6pt}\\ 
\parbox[t]{15cm}{
$^1$ Grupo de Investigaciones en Astronom\'{\i}a Te\'orica y Experimental, 
IATE, Observatorio Astron\'omico, Laprida 854, CP X5000BGR, 
C\'ordoba, Argentina \\
$^2$ Consejo Nacional de Investigaciones Cient\'{\i}ficas y T\'ecnicas 
de la Rep\'ublica Argentina, CONICET, Avenida Rivadavia 1917, CP C1033AAJ,
Buenos Aires, Argentina.
}
}
\date{\today}

\maketitle

\begin{abstract}
A comprehensive study on compactness has been carried out on the 2dF Galaxy
Group Catalogue constructed by Merch\'an \& Zandivarez. The compactness
indexes defined in this work take into account different geometrical 
constraints in order to explore a wide range of possibilities. 
Our results show that there is no clear distinction between groups
with high and low level of compactness when considering particular properties
as the radial velocity dispersion, the relative fraction of galaxies per 
spectral type and luminosity functions of their galaxy members.

Studying the trend of the fraction of galaxies per spectral type as a 
function of the dimensionless crossing time some signs of dynamical evolution 
are observed.
From the comparison with previous works on compactness we realize that special
care should be taken into account for some compactness criteria definitions
in order to avoid possible biases in the identification.
\end{abstract}


\section{Introduction} 
Compact groups (CGs) are small systems of a few 
galaxies which are in close proximity one another. 
Their are excellent laboratories for studying galaxy interactions
and, in particular, merging processes.
Given their high galaxy density (equivalent to those at the centers of
rich clusters) and small velocity dispersion ($\sim 200 \kms$), CG
members are expected to finally merge into one giant elliptical
galaxy within a few short crossing times.

Historically, CGs were of interest because of the obvious distortion 
of many of their member galaxies. 
The first systematic seek for CGs was pioneered by Rose (1977),
using a surface number density contrast procedure. 
The most widely analysed samples are the Hickson Compact Groups (HCGs) (Hickson 
1982, Hickson 1997), which have been selected on the basis of population, 
isolation (avoiding cores of rich clusters) and compactness.
Their compactness criteria involve the computation of the mean surface 
brightness of a group which should be lesser than a maximun limit. 
This mean was calculated distributing the flux of the member galaxies over 
the circular area containing their geometric centers. 
Previous samples have been visually selected, and thus reflect some of the 
systematic biases intrinsic to bidimensional identification of systems. 
A geometric bias arise because prolate systems along the line of sight will 
be preferentially selected. A kinematic bias could enhance the selection of 
systems which are in a transient compact configuration due to galaxy internal
motions.
Mamon (1986) has suggested that about half of HCGs are 
superpositions of galaxies within loose groups (hereafter LGs). 
This suggestion could imply that groups properties in any particular sample
may be strongly influenced by the criteria used to define the sample.
Another important clue in order to test the compact groups environment is
the color of their galaxy members. It is well known that elliptical 
galaxies recently formed from mergers of spiral galaxies should be bluer
than normal elliptical galaxies. This kind of interactions should be 
more frecuently observed in a compact group environment. Nervertheless,
a study on galaxy members of HCGs made by Zepf, Whitmore \& Levinson (1991)
have show that most of the early-type galaxies have optical colors 
indistinguishable form those of elliptical galaxies in other environments.
Furthermore, there is evidence of a strange absence of strong signs of
interactions, strong radio sources or far infrared radiation emission, etc.
(Menon 1995, Pildis, Bregman \& Schombert 1995, Sulentic 1997).
Recently, Tovmassian, Yam \& Tiersch (2001) and Tovmassian (2001) presents new 
evidence that indicates that almost all HCGs are dynamically associated 
generally with elongated LGs which are distributed along the elongation
of the corresponding groups, suggesting that the HCGs are the compact cores 
of the latter.
Consequently, an important question about the real nature of CGs arise: 
are CGs a distinct class by themselves or are extreme examples of systems 
having a particular range of galaxy density and population.

This question can be addressed studying the spatial distribution and 
environment of CGs. However, estimating the velocity dispersion and physical 
separations of galaxies in groups with a small number of galaxies is highly 
uncertain. Meaningful conclusions on dynamical properties about systems 
containing only four or five galaxies requires statistical analysis of large 
homogeneous samples.

\begin{figure}
\epsfxsize=0.5\textwidth
\hspace*{-0.5cm} \centerline{\epsffile{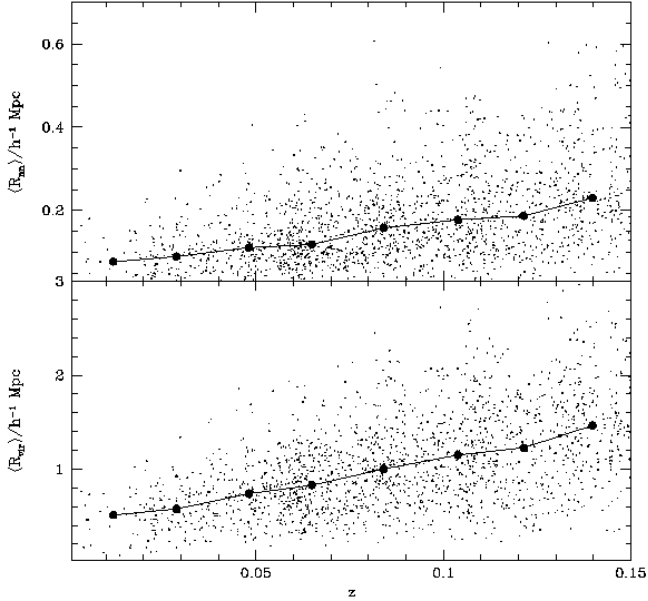}}
\caption{ 
The median mean nearest neighbour separation $\langle R_{nn}\rangle$ 
(upper panel) and the median virial radius $\langle R_{vir}\rangle$ 
(lower panel) as a function of redshift (filled circles) superimposed
to the real data (dots) distribution.
}
\label{fig0}
\end{figure}

\begin{figure}
\epsfxsize=0.5\textwidth
\hspace*{-0.5cm} \centerline{\epsffile{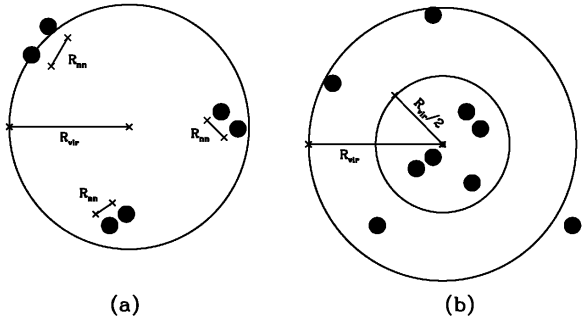}}
\caption{ 
Example (a): Possible configuration of binary galaxy members of a group
separated one another by large distances. Example (b): Configuration
of the group galaxy members showing the possibility to find a loose group
with a central concentration of galaxies.
}
\label{fig00}
\end{figure}

\begin{figure}
\epsfxsize=0.5\textwidth
\hspace*{-0.5cm} \centerline{\epsffile{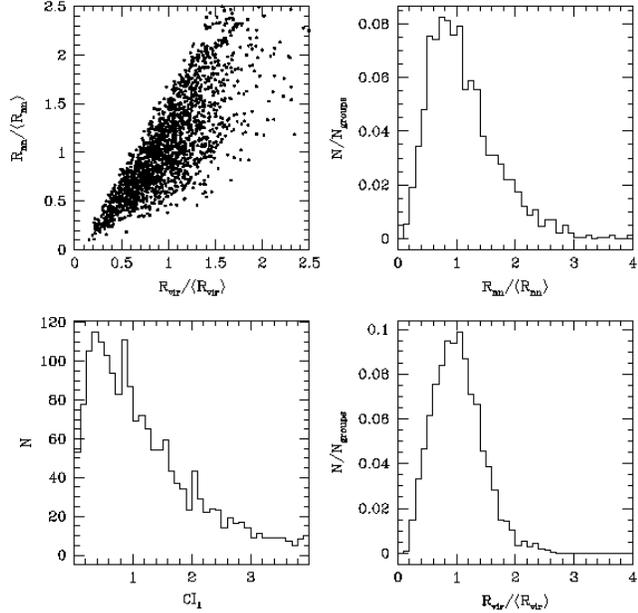}}
\caption{ 
Upper left panel: Scatter plot of the normalized mean nearest neighbour 
separation of group members in the 2dFGGC versus the normalized virial radius.
Upper right panel: Distribution of the normalized mean nearest neighbour 
separation.
Lower right panel: Distribution of the normalized virial radius.
Lower left panel: Distribution of the compactness index 1 ($CI_1$) defined in
section 3 for groups in the 2dFGGC. 
}
\label{fig1}
\end{figure}

In order to overcome these biases, it has recently become feasible to find 
CGs using automatic identification. 
Such a procedure has the advantage of generate a sample that is homogeneous 
and complete within the criteria specified for the search.
Barton et al. (1996) have used a selection criteria (friends-of-friends 
algorithm) based only on physical extent and association in redshift space. 
Eventhough Hickson's isolation criteria is not present at all in 
Barton's work, this technique is more effective in finding groups
in regions of higher galaxy density because foreground and background galaxies
are automatically eliminated by the velocity selection criteria.
Other automatic algorithm for the selection of CGs from large galaxy
catalogues has been developed by Iovino et al. (1999). 
The algorithm is such as to maximize the probability that the groups selected 
are physically related and partially reproduces the criteria used in the 
visual search by Hickson (1982), where his isolation criteria is slightly 
relaxed.\\
Since it is very difficult to identify compact groups at high redshifts,
more reliable results can be obtained restricting the analysis to low
redshift samples. Under this constraint applied on the Updated Zwicky Catalogue,
Focardi \& Kelm (2002) have shown that triplets are characterized by 
different properties than that obtained for higher order compact groups
suggesting the existence of two different galaxy systems in 
the compact group samples.
It is therefore of great interest to obtain larger and deeper samples of
CGs, in order to put the CGs properties on a statistical basis.
This will help to work out the controversy around the properties of CGs, 
contradictions that may only be apparent, given our still limited knowledge 
of the nature of these structures.

Currently, one of the largest group catalogues (hereafter 2dFGGC) was 
constructed by Merch\'an \& Zandivarez (2002).
They have identified groups in the 2dF public 100K data release using a 
modified Huchra \& Geller (1982) group finding algorithm that takes into 
account 2dF magnitude limit and redshift completeness masks.
This catalogue constitutes a large and suitable sample for the study of both,
processes in group environments and the properties of the group population 
itself. The global effects of group environment on star formation
was analysed by Mart\'{\i}nez et al (2002a) using this catalogue.
Dom\'{\i}nguez et al (2002) presented hints toward understanding local  
environment effects on the spectral types of galaxies in groups
by studying the relative fractions of different spectral types
as a function of the projected local galaxy density and
the group-centric distance. Recently, an extensive statistical analysis on
galaxy luminosity function in groups was carried out by Mart\'{\i}nez et al
(2002b). 

The aim of this work is to perform an analysis on the groups in the 2dFGGC 
by defining new compactness indexes which are assigned to every group 
in the sample. Several studies have been made
using galaxy spectral type, velocity dispersion, luminosity and crossing time
as a function of the compactness indexes.
The outline of this paper is as follows. In section 2, we present
the 2dF Galaxy Group Catalogue (2dFGGC) used throughout this work. 
Section 3 describes the compactness indexes definitions
while in section 4 we analyse the possible dependence of our indexes with
group and galaxy properties. Finally, in section 5 we summarize our 
conclusions.

\section{The 2dFGGC}

Samples of galaxies and groups used in this work are the same 
used by Mart\'{\i}nez et al. (2002b) which were constructed taking into account
a revised version of the masks and mask software of the 
2dFGRS 100k data release, which includes the $\mu$-masks described in 
Colless et al. (2001).
The group catalogue is obtained following the same procedure as described 
by Merch\'an \& Zandivarez (2002). 
The finder algorithm used for group identification is similar to that 
developed by Huchra \& Geller (1982) but modified in order to take into account 
redshift completeness, magnitude limit and the magnitude completeness 
mask ($\mu$-mask) present on the current release of galaxies. 
The revised group catalogue comprises a total number
of 2198 galaxy groups with at least 4 members and
mean radial velocities in the range $900 ~\kms\leq V \leq 75000~\kms$.
These groups have a mean velocity dispersion of 
$265~ \kms$, a mean virial mass of $9.1\times 10^{13} \ h^{-1} \ 
M_{\odot}$ and a mean virial radius of $1.15 ~ \mpc$.
Throughout this work we adopted the cosmological model $\Omega_0=0.3$ and
$\Omega_{\Lambda}=0.7$. 

\begin{figure}
\epsfxsize=0.5\textwidth
\hspace*{-0.5cm} \centerline{\epsffile{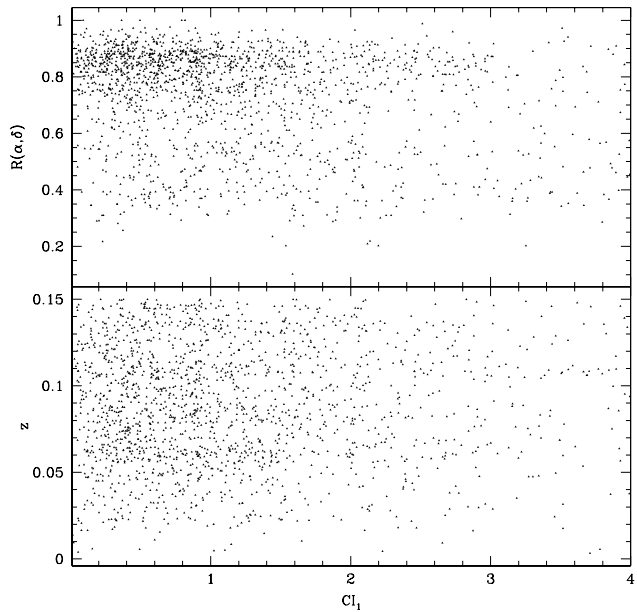}}
\caption{ 
Scatter plot of the $CI_1$ versus redshift completeness (upper panel)
and redshift (lower panel). 
}
\label{fig2}
\end{figure}

\begin{figure}
\epsfxsize=0.5\textwidth
\hspace*{-0.5cm} \centerline{\epsffile{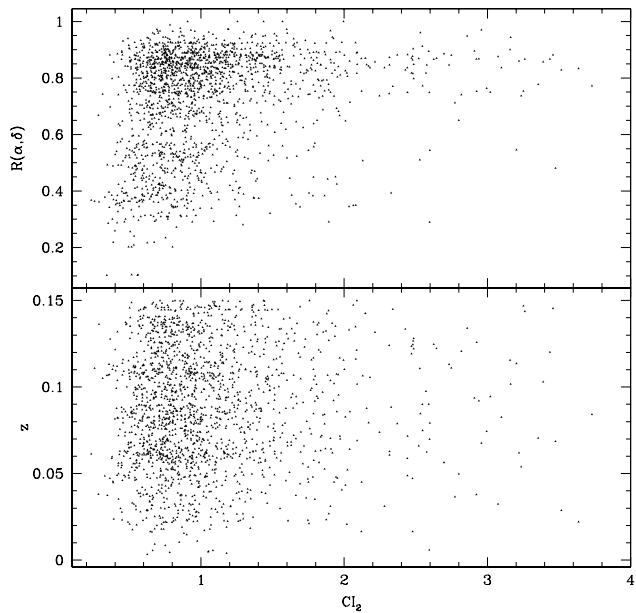}}
\caption{ 
Scatter plot of the $CI_2$ versus redshift completeness (upper panel)
and redshift (lower panel). 
}
\label{fig3}
\end{figure}

\section{Compactness index definitions}
The selection criteria used by Merch\'an \& Zandivarez (2002) based on
adaptable linking length parameters allow the identification of galaxy
systems independently of compactness.
We propose two quantities which measure the level of compactness for groups 
in the 2dFGGC. 
\begin{figure*}
\epsfxsize=0.7\textwidth
\hspace*{-0.5cm} \centerline{\epsffile{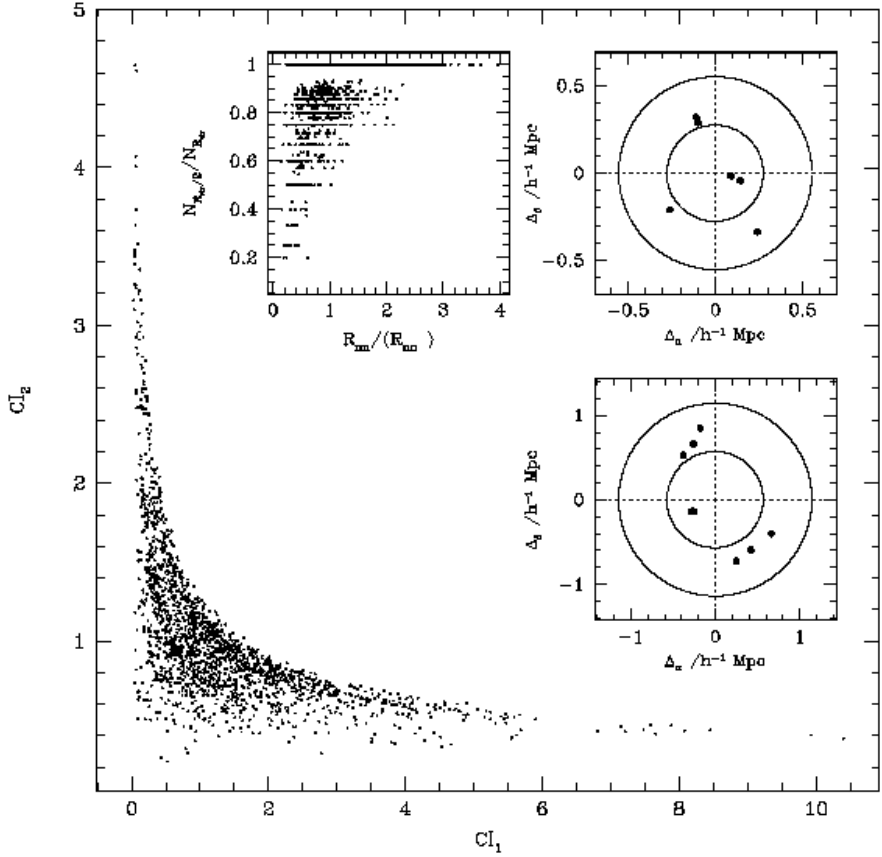}}
\caption{ 
Scatter plot of $CI_1$ versus $CI_2$. 
Inset upper left plot shows the scatter plot of $N_{R_{vir}/2}/N_{R_{vir}}$
versus $R_{nn}/\langle R_{nn} \rangle$. 
The upper and lower right panels show two
different groups projected on the sky with low values of $CI_1$ and 
$CI_2$. In these plots, the outer circle corresponds to the $R_{vir}$ and
the inner circle to $R_{vir}/2$.
}
\label{fig4}
\end{figure*}
These quantities are based on geometrical criteria and their application 
upon a group catalogue identified in redshift space provides a sample 
which is not constrained by the usual compactness selection criteria.
The first index is defined using the mean nearest projected neighbour 
separation of galaxies in groups $R_{nn}$. This quantity is normalized to the
mean of $R_{nn}$ for groups in a given bin of redshift $\langle R_{nn} 
\rangle (z)$ where the size of the redshift bin is $\sim 0.018$. 
This normalization is important in order to 
avoid a redshift dependence of this parameter since we are working with a 
magnitude limited sample; thus the mean nearest projected neighbour separation
increases its value with redshift (see upper panel of Figure \ref{fig0}). 
Nevertheless, this ratio is not enough to characterize the compactness of a 
group. For instance, a group formed by a set of binaries separated one another 
by large distances would have a small value of $R_{nn}$ (see example (a)
of Figure \ref{fig00}). Attempting to improve our definition
we introduce a virial radius $R_{vir}$ dependence. This parameter is also 
conveniently normalized to the redshift dependent mean virial radius 
$\langle R_{vir} \rangle(z)$. 
Median values, for redshift bin, of $R_{vir}$ as a function of $z$ are 
shown in the lower panel of Figure \ref{fig0}.
Consequently, the compactness index is defined by
\begin{equation}
CI_1= \frac{R_{nn}}{\langle R_{nn} \rangle (z)} \times  \frac{R_{vir}}{\langle R_{vir} \rangle (z)}
\end{equation}
In Figure \ref{fig1} we observe the distribution of both parameters 
involved in the $CI_1$ definition. Upper and lower right panels show
the normalized distributions of each ratio. 
The similarity of both distributions implies that neither of them is the 
dominant term in the compactness index definition.
Upper left panel of Figure \ref{fig1} shows the scatter plot for the two 
ratios, indicating the importance of using both parameters to define the 
compactness index.
The $CI_1$ distribution is plotted in the lower left panel of this figure. 
The groups related with a high level of compactness have the lowest values
of $CI_1$, whereas very loose groups belong to the tail of the distribution.
\begin{figure*}
\epsfxsize=1.0\textwidth
\hspace*{-0.4cm} \centerline{\epsffile{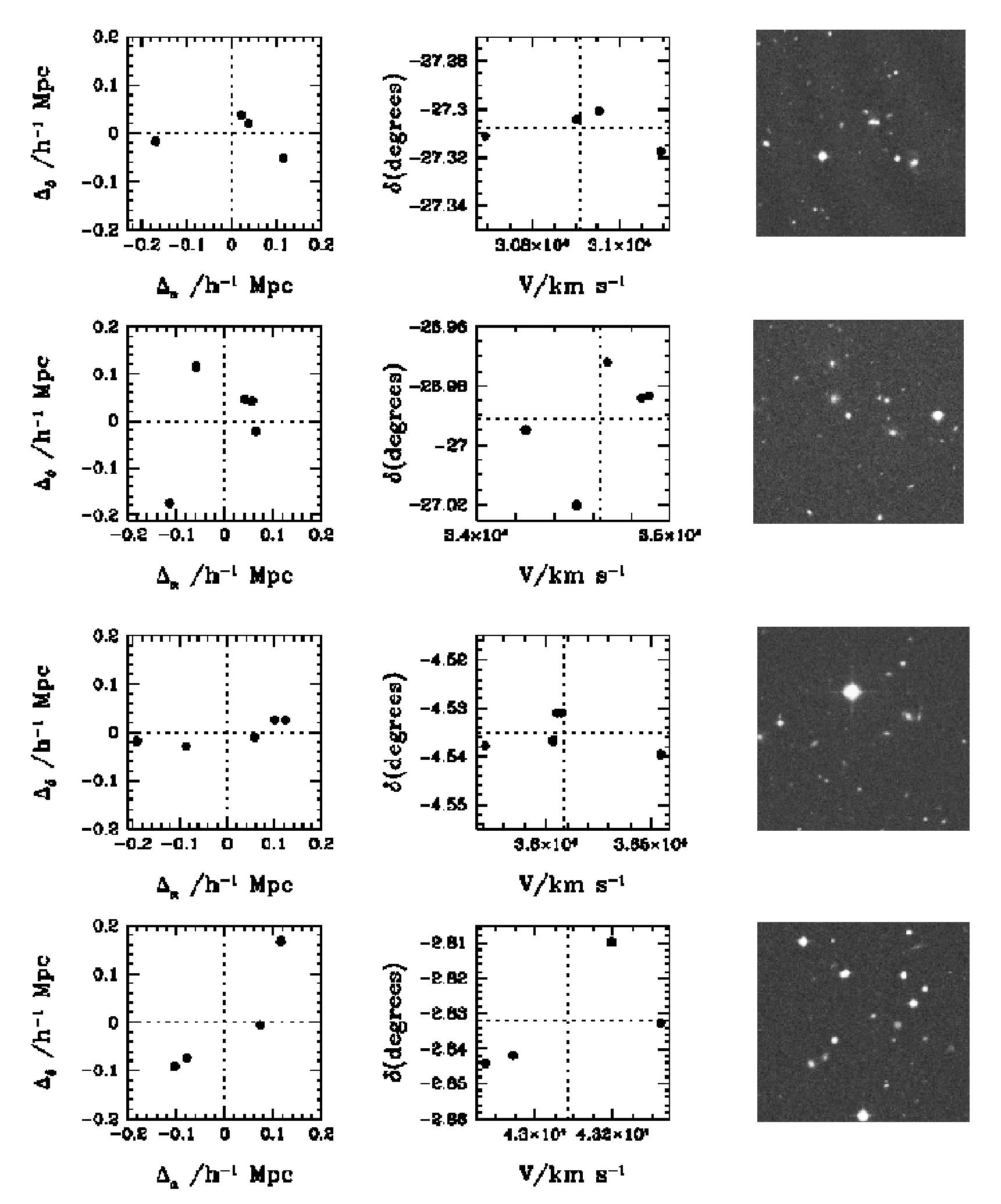}}
\caption{ 
Four examples of groups with high level of compactness for both indexes. 
The left panels shows the projection of equatorial coordinates in the sky. 
The distance units are in $h^{-1}\ Mpc$ and are measured from the group 
center of mass. The right panels shows the images taken from The
SuperCOSMOS Sky Survey (Hambly et al. 2001) of $4$ minutes of arc per side. 
The central panels shows the group projections in declination versus radial 
velocity. 
}
\label{fig5}
\end{figure*}
As stated above our compactness definition avoids any redshift dependence.
This fact is shown in Figure \ref{fig2} (lower panel) where an uniform
distribution is observed for the scatter plot of $CI_1$ versus $z$.
Given the status of the current release of the 2dF galaxy redshift survey,
it should be taken into account a possible dependence of $CI_1$ on 
the level of completeness of each group in the parent catalogue 
($R(\alpha,\delta)$). In the upper panel of Figure \ref{fig2} we display the
scatter plot of $R(\alpha,\delta)$ vs $CI_1$. This plot shows a total lack of 
correlation between the compactness index and the redshift completeness of
groups. Moreover, most of the groups show a high level of completeness
which provides confidence to our study.  

As mentioned in section 1, finding compact groups within loose groups is 
a possible scenario. In order to take into account this possibility, we 
define a new index which is able to identify this particularity. 
The second index measures the ratio of the number of galaxies enclosed 
in $R_{vir}/2$ to those within $R_{vir}$ (see example (b) of Figure 
\ref{fig00}). As before, this ratio is a 
measure of the core concentration regardless of the size of the system. 
Consequently, and keeping in mind that higher ratios imply higher central
concentration in groups, we divide this parameter by the normalized virial
ratio defined for index $CI_1$. Therefore, the second compactness index is 
\begin{equation}
CI_2= \frac{N_{R_{vir}/2}}{N_{R_{vir}}} \times  \Big(\frac{R_{vir}}{\langle R_{vir} \rangle (z)}\Big)^{-1}
\end{equation}   

A similar analysis as the one made for $CI_1$ shows that the 
second compactness index $CI_2$ depends neither on 
redshift nor on parent catalogue completeness (Figure \ref{fig3}). 
Furthermore, upper panel of Figure \ref{fig3} shows that most groups with high
level of compactness ($CI_2 \ge 2$) have the highest level of completeness.
To study a possible correlation between both indexes we show in Figure 
\ref{fig4} the scatter plot of $CI_1$ versus $CI_2$.
From this plot we observe that the groups in the lower left corner have
a high level of compactness for $CI_1$ whereas qualify as loose groups
according to $CI_2$. 
Two examples of groups with 6 and 8 galaxy members 
are shown in the inset right plots of Figure \ref{fig4}.
None of the examples shows a core concentration which implies a low $CI_2$ 
value, while the small mean nearest neighbour separations of galaxies
are the main responsible for the low values of $CI_1$.
These facts imply that a group with a high level of compactness for
$CI_2$ is also compact for $CI_1$. On the other hand small values for $CI_1$
do not guarantee high values of $CI_2$. Nevertheless, $CI_1$ is a better
discriminator of non central galaxy concentrations within a group. 
In the inset left plot of Figure \ref{fig4} we show the scatter plot of the 
compactness indexes without the normalized virial radius observing that 
the particular envelope in the main figure is due to this ratio.

As an example of groups characterized by a high level of compactness for
both indexes we show in Figure \ref{fig5} four of these systems.
The left panels show the projection of galaxy members on the sky, where the 
physical distances are referred to the center of mass of the system. 
These projections
can be directly correlated with the optical images shown in right panels.
Central panels display the declination-radial velocity projection of
groups. Previous plots show that the observed galaxy members are restricted
to a small range in redshift space ($\sim 800 \kms$).

\section{Analysing groups with $CI$}

\begin{figure}
\epsfxsize=0.5\textwidth
\hspace*{-0.5cm} \centerline{\epsffile{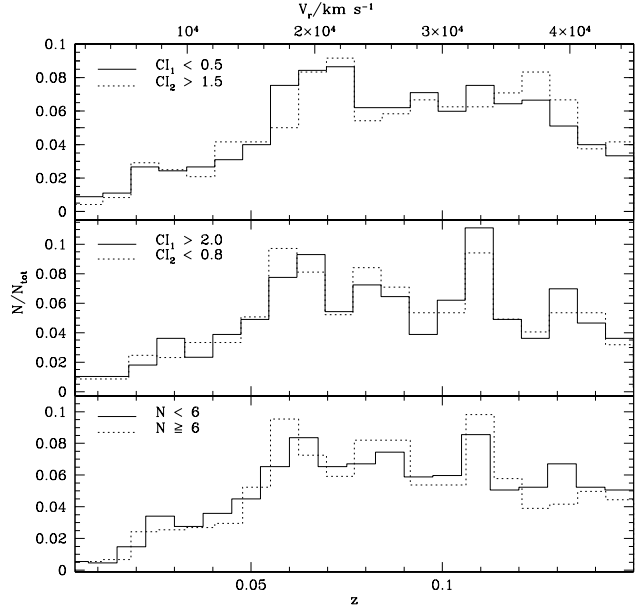}}
\caption{ 
Redshift normalized distribution of groups with high (upper panel) and
low (central panel) level of compactness for both indexes (see labels).
The lower panel shows the redshift normalized distribution of the group
sample depending on the number of galaxy members.
}
\label{fig60}
\end{figure}

\begin{figure}
\epsfxsize=0.5\textwidth
\hspace*{-0.5cm} \centerline{\epsffile{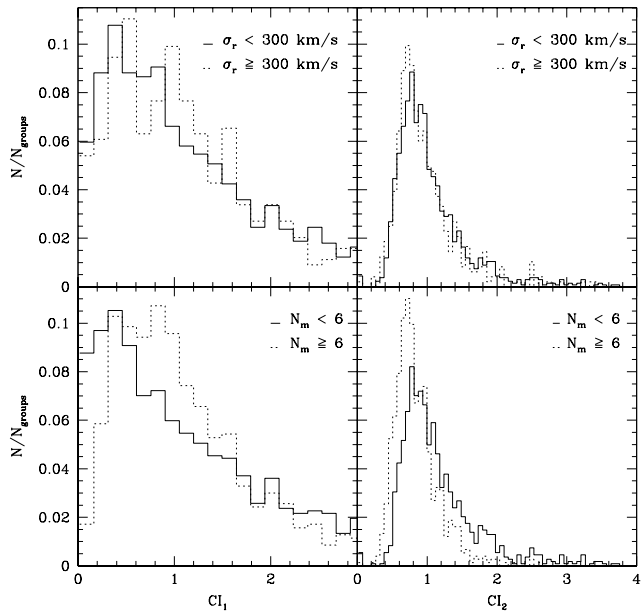}}
\caption{ 
Distribution of the compactness indexes for subsamples defined by velocity 
dispersion (upper panels) and number of members (lower panels).
}
\label{fig6}
\end{figure}

In the following analysis we study group and galaxy properties
for group subsamples defined by their compactness indexes. We
have chosen the following criteria:
\begin{itemize}
\item High level of compactness: $CI_1 < 0.5$ or $CI_2>1.5$
\item Low level of compactness: $CI_1 > 2.0$ or $CI_2<0.8$.
\end{itemize}
obtaining $451 \ (\sim 25\%)$ and $240 \ (\sim 13\%)$ groups with high level of 
compactness meanwhile the low level of compactness samples comprise 
$387 \ (\sim 21\%)$ and $690 \ (\sim 38\%)$ groups respectively. 
These limits were chosen to deal with the tails of the compactness index 
distributions while keeping an amount of groups large enough to
obtain a reliable statistics.

The redshift distributions of groups with low and
high level of compactness for both indexes are  
plotted in upper and central panels of Figure \ref{fig60}.
We observe quite similar redshift distributions for high and low
level of compactness irrespectively of the index used in the selection. 

\subsection{Radial velocity dispersion and galaxy members}

We look for any possible relation linking compactness indexes $CI_1$ and
$CI_2$, group velocity dispersion and group member richness.
In analysing radial velocity dispersion dependence we split the group sample 
in two subsamples with $\sigma_r < 300 \kms$ and $\sigma_r \ge 300 
\kms$ obtaining roughly the same number of groups in both subsamples. 
The obtained normalized continuous distributions for $CI_1$ and $CI_2$ are 
plotted in the upper panels of Figure \ref{fig6}.
From these plots we observe that velocity dispersion does not discriminate
levels of compactness of groups in the 2dFGGC.
The velocity dispersion limit is not a critical issue, we have 
obtained the same results varying the adopted cut-off from $\sigma_r=200 \kms$
to $400\kms$.

We use a similar criteria to analyse a possible dependence of group galaxy
members on compactness. In this case, we split the group sample in two 
subsamples with $N_m < 6$ and $N_m \ge 6$ and plot the normalized distributions
in the lower panels of Figure \ref{fig6}.
Both distributions show that groups with high level of compactness 
have typically a small number of members.
From our analysis it is not clear whether this is a consequence of our
definitions of compactness or a real physical phenomena.  
These results are not biased by any possible redshift dependence
as can be observed in the lower panel of Figure \ref{fig60}, where we plot
the redshift distribution of the subsamples defined by $N_m < 6$ and 
$N_m \ge 6$. We also observe that when increasing the galaxy member limit 
$N_m$, the results hold. Nevertheless by doing so the richer group sample 
becomes statistically poor.

\begin{figure}
\epsfxsize=0.5\textwidth
\hspace*{-0.5cm} \centerline{\epsffile{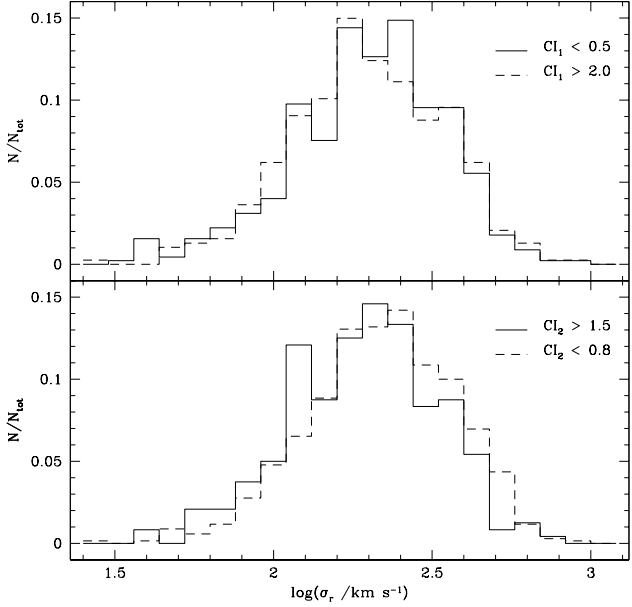}}
\caption{ 
Velocity dispersion distributions for subsamples with high and low level
of compactness. Upper panel corresponds to $CI_1$, whereas lower panel 
displays distributions for $CI_2$. The normalization factor $N_{tot}$
is the total number of groups in each subsample.
}
\label{fig7}
\end{figure}

As a complementary test, and using the samples defined at the beginning  
of this section, we plot the measured radial velocity dispersion
distributions. These distributions are shown in Figure \ref{fig7} 
where continuous line corresponds to the sample with high level of 
compactness while dashed line refers to the loose group sample.
It can be seen that the normalized distributions (in both panels) are quite 
similar showing a mean radial velocity dispersion of $\sim 200 \kms$.
This result is in agreement with the previously obtained by Hickson (1997),
where characteristic velocity dispersion of compact groups is quite similar
to that obtained for loose ones. This agreement is not obvious since our 
definition of compactness is very different to the one defined by Hickson.  

\subsection{Galaxy spectral type}

The following analysis is performed using the classification of galaxies 
defined by Madgwick et al. (2002) according to their spectral type.
This classification is based on the $\eta$ parameter which is very tightly
correlated with the equivalent width of $H_{\alpha}$ emission line,  
correlates well with morphology and can be interpreted
as a measure of the relative current star-formation present in each galaxy.
The four spectral types are defined as:
\begin{itemize}
\item Type 1: $~~~~~~~~~~\eta < -1.4$,
\item Type 2: $-1.4\leq \eta < ~~1.1$,
\item Type 3: $~~1.1 \leq \eta < ~~3.5$, 
\item Type 4: $~~~~~~~~~~\eta\ge ~~3.5$.
\end{itemize}
The Type 1 class is characterised by an old stellar population and
strong absorption features, the Types 2 and 3 comprise spiral
galaxies with increasing star formation, finally the Type 4
class is dominated by particularly active galaxies such as starburst.

Recently, using the 2dFGGC, Mart\'{\i}nez et al. (2002a) obtained a strong 
correlation between the relative fraction of galaxies
with high star formation (Type 4) and the parent group virial mass.
They found that even in the environment of groups with low virial masses
($M \sim 10^{13} \ M_{\odot}$) the star formation of their member galaxies
is significantly suppressed. 
Another study with relative fraction of galaxy spectral types in the 2dFGGC 
was performed by Dom\'{\i}nguez et al. (2002).
They found a clear distinction between high virial mass groups 
($M_V\gsim 10^{13.5} M_{\odot}$) and the less massive ones.
While the massive groups show a significant dependence of the relative 
fraction of low star formation galaxies on local galaxy density and 
group-centric radius, groups with lower masses show no significant trends.

\begin{figure}
\epsfxsize=0.5\textwidth
\hspace*{-0.5cm} \centerline{\epsffile{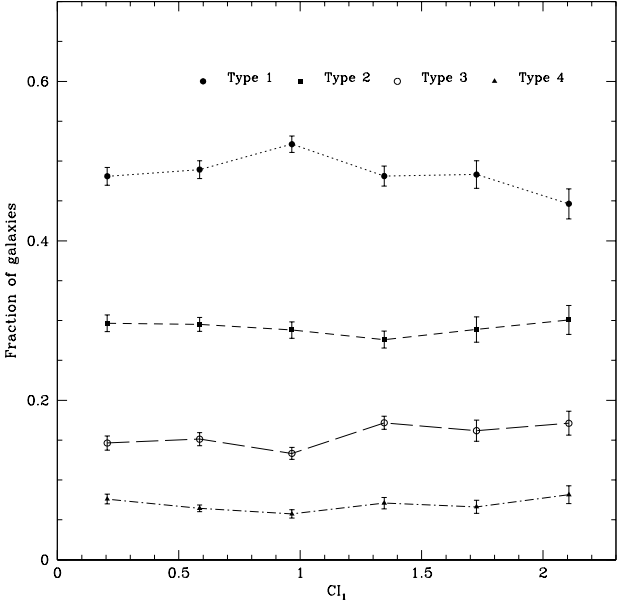}}
\caption{ 
Fraction of galaxies per spectral type as a function of $CI_1$. Error bars 
were estimated by the bootstrap resampling technique. Dotted line correspond
to Type 1 galaxies, short-dashed line to Type 2, long-dashed to Type 3 and
dot-dashed to Type 4.
}
\label{fig8}
\end{figure}

\begin{figure}
\epsfxsize=0.5\textwidth
\hspace*{-0.5cm} \centerline{\epsffile{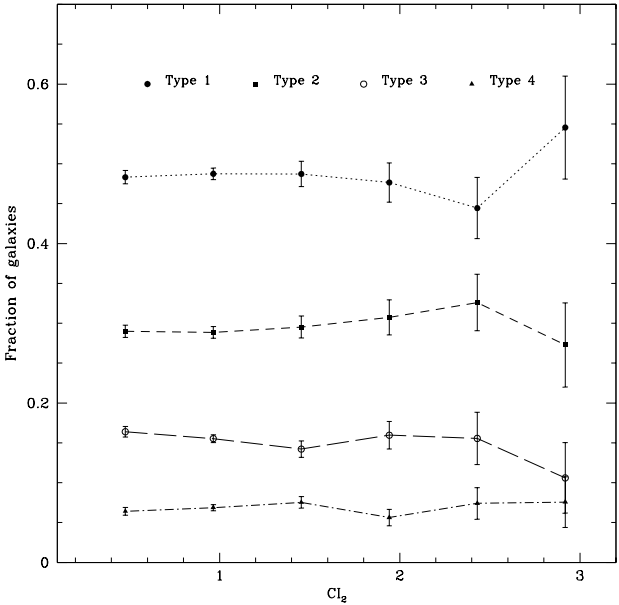}}
\caption{ 
Fraction of galaxies per spectral type as a function of $CI_2$. Error bars 
were estimated by the bootstrap resampling technique. Dotted line correspond
to Type 1 galaxies, short-dashed line to Type 2, long-dashed to Type 3 and
dot-dashed to Type 4.
}
\label{fig9}
\end{figure}

Following the study on possible dependences 
of the galaxy spectral type fractions on the environment,
we study here the behaviour of this fractions as a function of the previously 
defined compactness indexes $CI_1$ and $CI_2$.
In figure \ref{fig8} and \ref{fig9} we display galaxy fractions
per spectral type as a function of $CI_1$ and $CI_2$. Error bars 
were estimated by the bootstrap resampling technique. 
No statistically significant trends of the fraction of spectral types 
on the compactness are appreciated.

\subsection{Luminosity function of galaxies in groups}

A possible influence of compactness on the luminosity of galaxies 
could be studied using the luminosity functions (LF) of galaxies in
groups by defining two subsamples according to the level of compactness. 
The high level of compactness samples is defined by 
$CI_1 \le 0.5 $, $CI_2 \ge 1.2$, whereas the loose samples comprise groups
with $CI_1 \ge 2.0 $, $CI_2 \le 0.55$. These limits are chosen in order
to obtain samples with a large number of galaxies suitable for our computation.

In this work we use, the $C^-$ method (Lynden-Bell 1971) to make a non 
parametric determination of the LF. 
This method is the best estimator to measure the LF and is the less
affected by the faint end slope of a Schechter parametrisation or the 
sample size (Willmer 1997).
As a comparative rule we also use the STY (Sandage, Tammann \& Yahil 
1979) maximum likelihood Schechter fit to the LF of the whole group sample 
determined by Mart\'{\i}nez et al (2002b).
These authors found that the corresponding best fit Schechter parameters 
are $\alpha=-1.13\pm0.02$ and $M^{\ast}-5\log(h)=-19.90 \pm 0.03$, which 
are quite consistent with the results by Norberg et al. (2001) for field 
galaxies.

The adopted $C^-$ method estimator is the same as the used by Mart\'{\i}nez
et al. (2002b). This method is the version of Choloniewski (1987) developed
in an attempt to estimate both, the shape and normalization of the luminosity 
function. The LF is obtained by differentiating the cumulative LF, $\Psi(M)$. 
The function $X(M)$ defined as the observed density of galaxies with absolute 
magnitude brighter than $M$, represents only an undersampling of the $\Psi(M)$. 
Linden-Bell has defined a quantity $C(M)$ as the number of galaxies brighter 
than $M$ which could have been observed if their magnitude were $M$. This
quantity represents a subsample of $X(M)$ and compensates for the undersampling.
Taking into account the sky coverage of the 2dF present release, the 
differential LF can be written as
\begin{equation}
\langle \Phi(M) \rangle=\frac{\Gamma \ \sum_i^{M_i\in [M,M+\Delta M]}\psi_i}
{\Delta M}
\end{equation}
where
\begin{equation}
\Gamma=\prod_{k=2}^{N}\frac{C_k+w_k}{C_k}\left(V \sum_{i=1}^{N}\psi_i
\sum_{j=1}^{N}\frac{R(\alpha_j,\delta_j)}{N}\right)^{-1},
\end{equation}
\begin{equation}
\psi_k=\prod_{i=1}^{k}\frac{C_i+w_i}{C_{i+1}}
\end{equation}
and $R(\alpha,\delta)$ is the redshift completeness of the parent catalogue.
$C_k \equiv C^{-}(M_k)$ is defined as $C(M)$ but excluding the object
$k$ itself and weighting each object by $w$ which is the inverse of the 
magnitude-dependent redshift completeness (Norberg et al. 2002, Mart\'{\i}nez
et al. 2002b).

\begin{figure}
\epsfxsize=0.5\textwidth
\hspace*{-0.5cm} \centerline{\epsffile{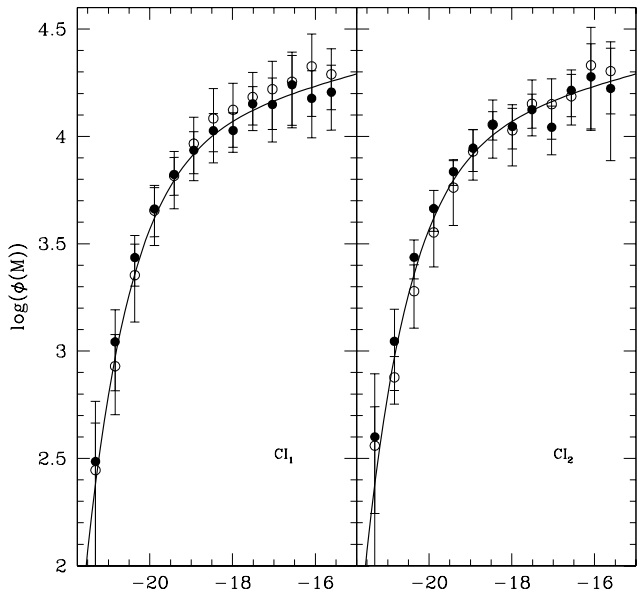}}
\caption{ 
Luminosity functions of galaxies in groups. The groups are separated in
high (filled circles) and low (open circles) compactness using 
the $CI_1$ (left panel) and $CI_2$ (right panel).
The solid line shows the STY fit obtained by Mart\'{\i}nez et al. (2002b)
for the whole sample of galaxies in groups.
}
\label{fig10}
\end{figure}

In Figure \ref{fig10} we show the luminosity functions for
galaxies in groups in arbitrary units. 
Absolute magnitudes are computed using the $k+e$ mean correction for
galaxies in the 2dFGRS as derived by Norberg et al. (2002).
Error bars are estimated using 10 mock catalogues constructed from numerical
simulations of a cold dark matter universe according to the cosmological model
adopted in this work with a Hubble constant $h=0.7$ and a relative mass
fluctuation $\sigma_8=0.9$. These simulations were performed using $128^3$
particles in a cubic comoving volume of $180 \mpc$ per side.
In the left panel of Figure \ref{fig10}, filled circles correspond to the
LF of galaxies in groups with high level of compactness using $CI_1$,
while the open circles represent the LF for the very loose groups with the
same index. The right panel shows the analogous LF using $CI_2$ as
compactness discriminator. 
In both panels, solid line shows the STY best fit to the LF for the overall 
sample of galaxies in groups obtained by Mart\'{\i}nez et al. (2002b) 
which corresponds to the Schechter parameters $M^{\ast}=-19.90\pm 0.03$ and 
$\alpha=-1.13\pm 0.02$.
From this figure we observe that the resulting LF's are insensitive to the
level of compactness of groups when comparing with the whole sample.

\subsection{The dimensionless crossing time}

In order to consider a possible level of dynamical evolution of galaxy systems
we compute the dimensionless crossing time as used by Hickson et al. (1992). 
This particular timescale, $H_0 t_c$, is the ratio of the crossing time 
to the approximate age of the universe, and is defined by
\begin{equation}
H_0 t_c=\frac{(4 \times 100)}{\pi}\times \frac{\Delta}{\sigma}
\end{equation}
where $\Delta$ is the mean projected galaxy separation
in groups and $\sigma$ is the 3-dimensional velocity dispersion
($\sigma=\sqrt{3}\sigma_r$).
The dimensionless crossing time may reflect dynamical evolution 
since its inverse is roughly the maximun number of times that a galaxy could 
have traversed the group since its formation.   

Computing the crossing time for group having $CI_2 \ge 1.5$
we obtain the range $0.017 \ H_0^{-1}$ 
to $0.25 \ H_0^{-1}$ with a mean value of $\sim 0.09$. 
This value is significantly higher than the one obtained by Hickson et 
al. (1992).
Their compact group sample shows typically smaller crossing 
time values mainly due to small values of the mean galaxy-galaxy 
separation $\Delta$. A possible explanation could arise from the suggestion 
that many compact groups could be the cores of larger ones. 
Many studies have been carried out about this possibility
(Mammon 1986, Tovmassian et al. 2001) obtaining that most of the groups
identified as compact are the central region of larger groups.
This kind of misidentification could be a very important source of
bias in the study of compact groups.
In order to quantify this problem we compute the fraction of groups with
high central concentrations within larger groups using our catalogue. 
Seeking for groups with ratio $N_{R_{vir}/2}/N_{R_{vir}} \ge 0.9$ in 
larger groups we obtain that roughly the $70 \%$ of these groups are 
characterised by a high central concentration. 
Consequently, this kind of bias can be the explanation for the 
small crossing time values obtained for Hickson's sample.

\begin{figure}
\epsfxsize=0.5\textwidth
\hspace*{-0.5cm} \centerline{\epsffile{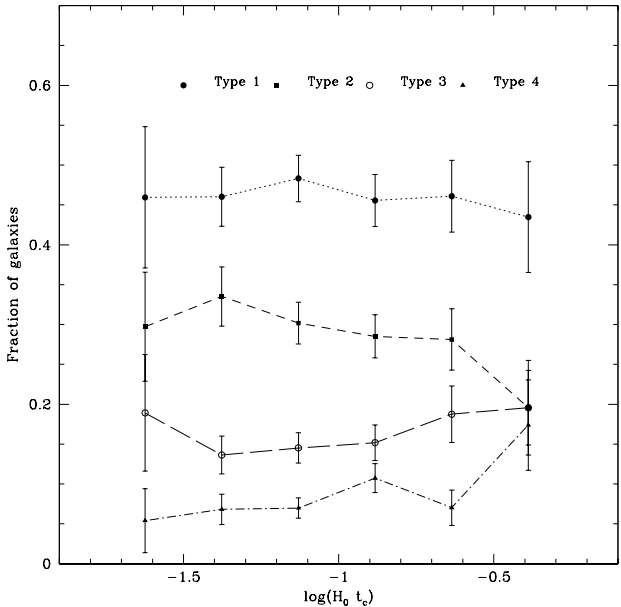}}
\caption{ 
Fraction of galaxies per spectral type as a function of the dimensionless
crossing time. 
}
\label{fig12}
\end{figure}

Finally, we intended to search for a possible dependence of the fraction
of galaxies per spectral type on the dimensionless crossing time.
Figure \ref{fig12} shows the trends obtained for the four spectral types 
defined in subsection 4.2. 
A slight increase of the Type 4 fraction of galaxies with the $H_0 t_c$ 
can be observed, meanwhile an opposite trend is shown by Type 2 galaxies.
Even when this result is not highly significant, a possible dynamical
evolution is suggested. Such evolution would consist in the conversion of 
late type galaxies to earlier ones by dynamical processes such as mergers.
More significant results have been obtained by Dom\'{\i}nguez et al. (2002)
analysing the spectral type vs local galaxy density relation specially
for subsamples of groups with high virial masses using the same catalogue.
We should remark that these results are almost independent of the level of 
compactness of galaxy groups. A similar analysis developed for a sample of 
nearby compact groups by Focardi \& Kelm (2002) have also shown signs of 
evolution of spectral content in the relation morphology-velocity dispersion 
mainly for low multiplicity compact groups (see Figure 9 in their work).

\section{DISCUSSION AND CONCLUSIONS}
In this work we report a statistical compactness analysis using the 
catalogue of galaxy groups constructed by Merch\'an \& Zandivarez (2002).
Given the size of the original sample and 3-dimensional information 
our study intend to characterize groups of galaxies according to the
level of compactness and analyse its influence on groups and galaxy members. 
For this purpose, we define two new compactness indexes
based on geometrical criteria. 
Whereas index $CI_1$ prioritizes the distance to the nearest neighbour and
the size of the system (eq. 1), index $CI_2$ enhances a possible core
concentration in the system (eq. 2).
Special cares have been taken over the construction of these indexes in order
to avoid possible dependences on redshift and redshift completeness of 
the parent catalogue (Figures \ref{fig2} and \ref{fig3}).

With this characterization, we develop a wide analysis over many
physical properties of groups and galaxy members.
First, we observe that the compactness indexes distribution shows 
identical behaviours when the group sample is split for low 
($\sigma_r < 300 \kms$) and high ($\sigma_r \ge 300 \kms$)
velocity dispersion. Furthermore, we observe that groups with high and low 
levels of compactness show the same normalized radial velocity dispersion 
distributions with a mean of $\sim 200 \kms$.
This result is consistent with previous ones which
reflect that the mean velocity dispersion of compact groups is quite
similar to that found for loose groups (Hickson et al. 1992).
Analysing the dependence of group compactness with numerical richness 
we observe that groups with high level of compactness have 
typically a low number of galaxy members, while most of the loose groups
are characterized by a larger number of galaxy members.

On the other hand, another study has been made about the fraction of galaxies 
per spectral type and luminosity as a function of the two compactness indexes
(Figures \ref{fig8}, \ref{fig9} and \ref{fig10}).
Our results do not show any particular correlation between the above 
parameters and the compactness level for groups in the sample. 

The similar behaviour observed in groups with high level of compactness
and loose ones probably suggests that this distinction is not fundamental.
This result is supported by previous works which state an 
indistinguishable behaviour between compact and loose groups showing
that many compact groups are located within overdense environments
(de Carvalho et al. 1997, Barton et al. 1998, Zabludoff \& Mulchaey 1998). 
Furthermore, an analysis on X-ray properties of groups shows that it is 
impossible to separate loose and compact groups on the luminosity-temperature
relation, the luminosity-velocity dispersion relation or in the velocity
dispersion-temperature relation stating that a more useful distinction is
that between X-ray bright and X-ray faint systems (Helsdon \& Ponman 2000).

The mean dimensionless crossing times obtained for a sample with high level of
compactness is shifted toward higher values when comparing to the obtained 
for a sample of compact groups constructed by Hickson et al. (1992). 
This shift could be due to some biases in the compact group identification 
criteria. These biases could imply the detection of the cores of larger 
systems generating smaller dimensionless crossing times determinations.

The last correlation we studied is the fraction of
galaxies per spectral type as a function of the dimensionless crossing time.
Eventhough the correlations we found are not significant, it is worth to mention
that for Type 2 galaxies, smaller is the fraction when higher is the 
dimensionless crossing time and the opposite trend is maintained for Type 
4 galaxies. 
This latter result is consistent with the stated by Hickson et al.
(1992) about groups with smaller crossing times typically containing
fewer late-type galaxies.

\section*{Acknowledgments}
We thank the referee Paola Focardi for helpful comments and suggestions
and Mirta Mosconi for carefully reading the manuscript.
We thank to Peder Norberg and Shaun Cole for kindly providing the 
software describing the mask of the 2dFGRS and to the 2dFGRS Team
for having made available the current data sets of the sample.
AZ and HJM are supported by fellowships from CONICET, Argentina.
MJLD is supported by a fellowship from SECyT, Universidad Nacional de 
C\'ordoba, Argentina. 
We also thank to J. Hetfield and L. Ulrich for support. 
This work has been partially supported by grants from 
the Secretar\'{\i}a de Ciencia y T\'ecnica de la Universidad Nacional de 
C\'ordoba (SeCyT) and Agencia C\'ordoba Ciencia.
This work made use of the C4JA facilities at IATE.

\end{document}